\PassOptionsToPackage{unicode}{hyperref}
\PassOptionsToPackage{hyphens}{url}
\PassOptionsToPackage{dvipsnames,svgnames,x11names}{xcolor}
\documentclass[
]{article}
\usepackage{amsmath,amssymb}
\usepackage{lmodern}
\usepackage{iftex}
\ifPDFTeX
  \usepackage[T1]{fontenc}
  \usepackage[utf8]{inputenc}
  \usepackage{textcomp} % provide euro and other symbols
\else % if luatex or xetex
  \usepackage{unicode-math}
  \defaultfontfeatures{Scale=MatchLowercase}
  \defaultfontfeatures[\rmfamily]{Ligatures=TeX,Scale=1}
\fi
\IfFileExists{upquote.sty}{\usepackage{upquote}}{}
\IfFileExists{microtype.sty}{% use microtype if available
  \usepackage[]{microtype}
  \UseMicrotypeSet[protrusion]{basicmath} % disable protrusion for tt fonts
}{}
\makeatletter
\@ifundefined{KOMAClassName}{% if non-KOMA class
  \IfFileExists{parskip.sty}{%
    \usepackage{parskip}
  }{% else
    \setlength{\parindent}{0pt}
    \setlength{\parskip}{6pt plus 2pt minus 1pt}}
}{% if KOMA class
  \KOMAoptions{parskip=half}}
\makeatother
\usepackage{xcolor}
\usepackage{longtable,booktabs,array}
\usepackage{calc} % for calculating minipage widths
\usepackage{etoolbox}
\makeatletter
\patchcmd\longtable{\par}{\if@noskipsec\mbox{}\fi\par}{}{}
\makeatother
\IfFileExists{footnotehyper.sty}{\usepackage{footnotehyper}}{\usepackage{footnote}}
\makesavenoteenv{longtable}
\usepackage{graphicx}
\makeatletter
\def\maxwidth{\ifdim\Gin@nat@width>\linewidth\linewidth\else\Gin@nat@width\fi}
\def\maxheight{\ifdim\Gin@nat@height>\textheight\textheight\else\Gin@nat@height\fi}
\makeatother
\setkeys{Gin}{width=\maxwidth,height=\maxheight,keepaspectratio}
\makeatletter
\def\fps@figure{htbp}
\makeatother
\NewDocumentCommand\citeproctext{}{}
\NewDocumentCommand\citeproc{mm}{%
  \begingroup\def\citeproctext{#2}\cite{#1}\endgroup}
\makeatletter
 \let\@cite@ofmt\@firstofone
 \def\@biblabel#1{}
 \def\@cite#1#2{{#1\if@tempswa , #2\fi}}
\makeatother
\newlength{\cslhangindent}
\newlength{\csllabelwidth}
\newenvironment{CSLReferences}[2] % #1 hanging-indent, #2 entry-spacing
 {\begin{list}{}{%
  \setlength{\itemindent}{0pt}
  \setlength{\leftmargin}{0pt}
  \setlength{\parsep}{0pt}
  \ifodd #1
   \setlength{\leftmargin}{\cslhangindent}
   \setlength{\itemindent}{-1\cslhangindent}
  \fi
  \setlength{\itemsep}{#2\baselineskip}}}
 {\end{list}}
\usepackage{calc}

\ifLuaTeX
\usepackage[bidi=basic]{babel}
\else
\usepackage[bidi=default]{babel}
\fi
\babelprovide[main,import]{american}
\def\languageshorthands#1{}
\ifLuaTeX
  \usepackage{selnolig}  % disable illegal ligatures
\fi
\IfFileExists{bookmark.sty}{\usepackage{bookmark}}{\usepackage{hyperref}}
\IfFileExists{xurl.sty}{\usepackage{xurl}}{} % add URL line breaks if available
\hypersetup{
  pdftitle={TriPoDPy: 1D Tri-Population size distributions for Dust
evolution in protoplanetary disks},
  pdfauthor={Nicolas Leo Kaufmann, Thomas Pfeil, Sebastian Stammler,
Anna B. T. Penzlin, Sandro Christian Paetzold, Til Birnstiel},
  pdflang={en-US},
  colorlinks=true,
  linkcolor={Maroon},
  filecolor={Maroon},
  citecolor={Blue},
  urlcolor={Blue},
  pdfcreator={LaTeX via pandoc}}

\title{TriPoDPy: 1D Tri-Population size distributions for Dust evolution
in protoplanetary disks}

\definecolor{c53baa1}{RGB}{83,186,161}
\definecolor{c202826}{RGB}{32,40,38}

\usepackage[affil-it]{authblk}
\usepackage{orcidlink}
\author[1%
  ]{Nicolas Leo Kaufmann%
    \,\orcidlink{0009-0005-4389-2947}\,%
    }
\author[2%
  ]{Thomas Pfeil%
    \,\orcidlink{0000-0002-4171-7302}\,%
    }
\author[1%
  ]{Sebastian Stammler%
    \,\orcidlink{0000-0002-1589-1796}\,%
    }
\author[1%
  ]{Anna B. T. Penzlin%
    \,\orcidlink{0000-0002-8873-6826}\,%
    }
\author[1%
  ]{Sandro Christian Paetzold%
    \,\orcidlink{0009-0005-5501-3620}\,%
    }
\author[1,3%
  ]{Til Birnstiel%
    \,\orcidlink{0000-0002-1899-8783}\,%
    }

\affil[1]{University Observatory, Faculty of Physics,
Ludwig-Maximilians-Universität München, Scheinerstr. 1, 81679 Munich,
Germany%
  }
\affil[2]{Center for Computational Astrophysics, Flatiron Institute, 162
Fifth Avenue, New York, NY 10010, USA%
  }
\affil[3]{Exzellenzcluster ORIGINS, Boltzmannstr. 2, D-85748 Garching,
Germany%
  }
\date{20 November 2025}

\begin{document}
\maketitle

\section{Summary}\label{summary}

\texttt{TriPoDPy} is a code simulating the dust evolution, including
dust growth and dynamics in protoplanetary disks using the parametric
dust model presented in (\citeproc{ref-Pfeil2024}{Pfeil et al., 2024}).
The simulation evolves a dust distribution in a one-dimensional grid in
the radial direction. It's written in \texttt{Python} and the core
routines are implemented in \texttt{Fortran90}. The code not only solves
for the evolution of the dust but also the gas disk with the canonical
\(\alpha\)-description (\citeproc{ref-Shakura1973}{Shakura \& Sunyaev,
1973}). In addition to the original model, we added descriptions of
tracers for the dust and gas, which could be used for compositional
tracking of additional components. The code is open source and available at \url{https://github.com/tripod-code/tripodpy}.

\section{Statement of Need}\label{statement-of-need}

Simulating the dust evolution in protoplanetary disks, including growth
and transport, is vital to understanding planet formation and the
structure of protoplanetary disks. There exist multiple open-source
codes that tackle this problem by either solving the Smoluchowski
Equation, e.g.~\texttt{Dustpy}(\citeproc{ref-Dustpy}{Stammler \&
Birnstiel, 2022}) or \texttt{CuDisc}(\citeproc{ref-CuDisc}{Robinson et
al., 2024}) or using a Monte Carlo approach (e.g.~\texttt{Mcdust}
(\citeproc{ref-Vaikundaraman}{Vaikundaraman et al., 2025})) to simulate
the mutual collisions between dust grains. However, all these
simulations are computationally expensive, which calls for parametrised
dust evolution models that can be used, for example, for population
studies. Previous models, e.g.~\texttt{Twopoppy}
(\citeproc{ref-twopop}{Birnstiel et al., 2012}), were not designed for
disks with radial sub-structures and were not calibrated for different
stellar masses.

These shortcomings are solved with the Tripod Dust model. It describes
the dust size distribution with a truncated power law, which allows the
simulation full access to the dust size distribution, which is essential
to accurately model the dust evolution and additional physical effects
like photoevaporation. Additionally, TriPodPy enables the addition of
tracers in gas and dust, which could be used for tracking of chemical
composition, electrical charge, and other parameters.

\section{Comparison Simulation}\label{comparison-simulation}

We compare a Simulation with our code with one performed with the full
coagulation code \texttt{Dustpy}, illustrating how well our code
performs. The parameters used for the comparison simulations can be
found in the Table below:

\begin{longtable}[]{@{}lc@{}}
\toprule\noalign{}
Parameter & Value \\
\midrule\noalign{}
\endhead
\bottomrule\noalign{}
\endlastfoot
gas surface density at 1 AU & 722 g/\(\mathrm{cm}^2\) \\
temperature at 1 AU & 209 K \\
turbulence strength (\(\alpha\)) & \(10^{-3}\) \\
fragmentation velocity (\(v_{\mathrm{frag}}\)) & 10 m/s \\
gas surface density power law \(p\) & 0.85 \\
temperature power law \(q\) & 0.5 \\
gap position & 5.2 AU \\
\(M_\text{planet}/M_{\star}\) & \(10^{-3}\) \\
\end{longtable}

We compare the particle size distribution from both simulations at
400'000 years, which can be seen in the figures below. The first figure
shows the dust surface density as a function of size and radius
throughout the disk (top) and the total dust surface desity as a
function of radius (bottom). The second plot shows 1-D slices at
different radii as indicated by the white dashed lines.

\includegraphics[keepaspectratio]{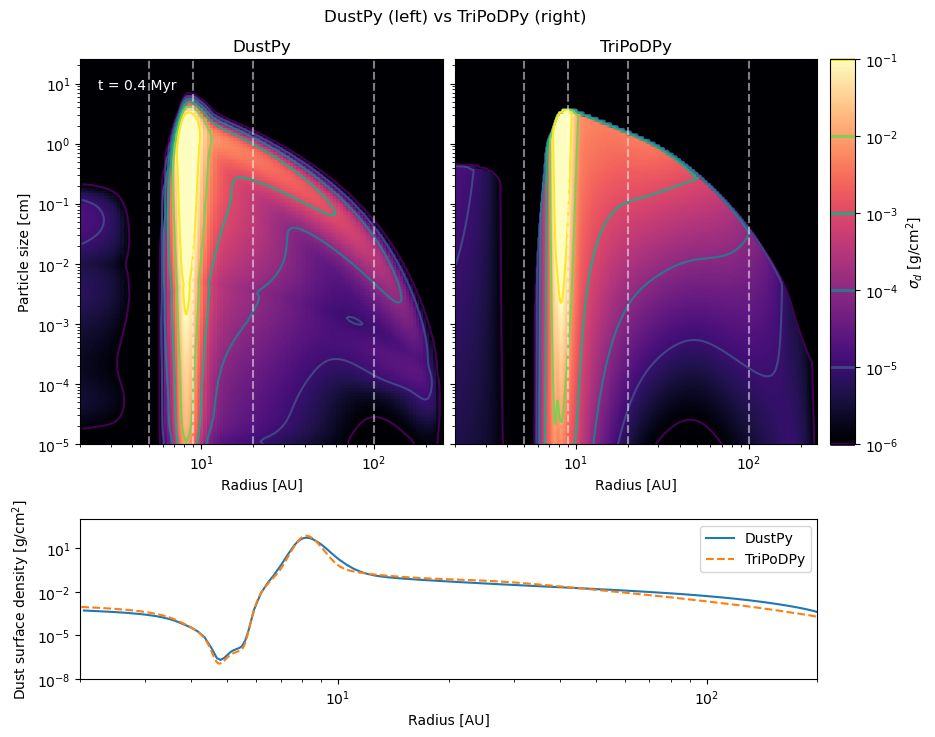}
\includegraphics[keepaspectratio]{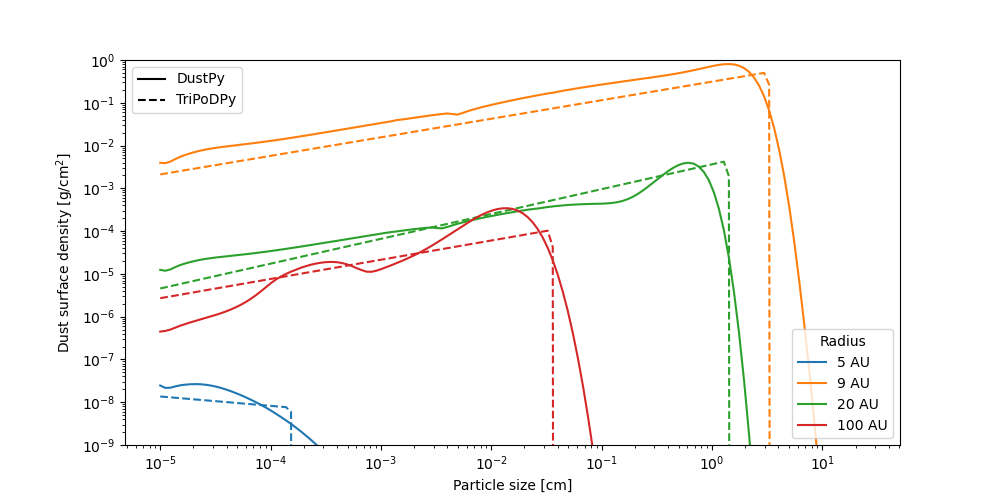}

The \texttt{TriPoDPy} simulation runs a factor of 50 to 100 faster than
the compared \texttt{DustPy} model. As we can see, the maximal sizes and
dust size distributions match quite well with the full coagulation code.
Since the size distribution is always assumed to be a power law,
capturing multimodal distributions is not possible, as can be seen
around 100 AU in the test simulation. This also affects the dust
distribution on the inside of the gap, as the dust size distribution in
gaps deviates from the expected power law as well. For an in-depth
discussion, see (\citeproc{ref-Pfeil2024}{Pfeil et al., 2024}).

\section{Acknowledgments}\label{acknowledgments}

The authors acknowledge funding from the European Union under the
European Union's Horizon Europe Research and Innovation Programme
101124282 (EARLYBIRD) and funding by the Deutsche Forschungsgemeinschaft
(DFG, German Research Foundation) under Germany's Excellence Strategy -
EXC-2094 - 390783311. Views and opinions expressed are, however, those
of the authors only and do not necessarily reflect those of the European
Union or the European Research Council. Neither the European Union nor
the granting authority can be held responsible for them. \# References

\phantomsection\label{refs}

\end{document}